\documentclass[a4paper]{article}
\usepackage[numbers]{natbib}

\usepackage[utf8]{inputenc}

\usepackage{graphicx}
\usepackage{amsmath}
\usepackage{multicol}
\usepackage[per-mode=symbol,]{siunitx}[=v2]
\usepackage{longtable,tabularx}
\usepackage{paralist, cleveref}
\usepackage{subcaption}

\newcommand{\kindex}[2]{\ensuremath{{#1}_{\scalebox{0.6}{#2}}}}

\usepackage{xspace}
\newcommand{\U}{\textrm{U}}
\newcommand{\Uinf}{\kindex{\U}{$\infty$}\xspace}
\newcommand{\Ueff}{\kindex{\U}{eff}\xspace}
\newcommand{\Ueffbar}{\kindex{\overline{\U}}{eff}\xspace}
\newcommand{\Uosc}{\kindex{\U}{osc}\xspace}

\newcommand{\Rey}{\textrm{Re}}
\newcommand{\ii}{\textrm{i}}

\graphicspath{{../../figurematter/plots/}{../../figurematter/latex/figs/}}

\title{Greenberg's force prediction for vertical-axis wind turbine blades}

\usepackage{authblk}
\author[1]{David Bensason}
\author[1]{Sébastien Le Fouest}
\author[2]{Anna Young}
\author[1]{Karen Mulleners\footnote{Corresponding author: karen.mulleners@epfl.ch}}
\affil[1]{Unsteady flow diagnostics laboratory, Institute of mechanical engineering, École Polytechnique Fédérale de Lausanne, 1015 Lausanne, Switzerland.}
\affil[2]{University of Bath, Bath, Somerset, BA2 7AY, United Kingdom}
\date{}                     
\begin{document}

\maketitle


\section*{Nomenclature}
{\renewcommand\arraystretch{1.0}
\noindent\begin{longtable*}{@{}l @{\quad=\quad} l@{}}
\kindex{a}{n} & $n^{\rm th}$ Fourier sine coefficient describing the effective angle of attack evolution\\
\kindex{b}{n} & $n^{\rm th}$ Fourier cosine coefficient describing the effective flow velocity evolution\\
$c$ & airfoil chord length, \si{\meter} \\
$C(k)=F(k)+\ii G(k)$ & Theodorsen's complex transfer function \\
$\kindex{C}{R}$   & radial load coefficient \\
$\kindex{C}{$\theta$}$   & tangential load coefficient \\
$d$ & chord-normalised distance between pitching axis and airfoil center\\
$h$ & vertical heaving displacement, \si{\meter}\\
$k=(\omega c)/(2u)$ & reduced frequency  \\
$P$ & total force on the airfoil, \si{\newton}\\
$R$ & turbine radius, \si{\meter} \\
$t$ & time, \si{\second} \\
$T$ & period for one turbine blade rotation, \si{\second} \\
$u$ & velocity encountered by airfoil, \si{\meter\per\second}\\
$\kindex{u}{0}$ & mean free-stream velocity, \si{\meter\per\second}\\
$\kindex{u}{0}\sigma$, $\kindex{\alpha}{0}$, $\kindex{h}{0}$ & surging, pitching, heaving amplitudes\\
$\kindex{\U}{osc}$ & oscillating contribution of effective flow velocity, \si{\meter\per\second}\\
$\Uinf$ & free stream velocity, \si{\meter\per\second} \\
$\Ueff$ & effective flow velocity, \si{\meter\per\second} \\
$\alpha$ & angle of attack, \si{\degree} \\
$\overline{\alpha}$ & average or fixed part of the angle of attack, \si{\degree} \\
$\kindex{\alpha}{eff}$  & effective angle of attack, \si{\degree} \\
$\kindex{\alpha}{eff}',\kindex{\U}{eff}'$ & Fourier series approximation of effective angle of attack and flow velocity \\
$\kindex{\alpha}{ss}$ & static stall angle, \si{\degree} \\
$\lambda$   & tip-speed ratio \\
$\rho$ & fluid density, \si{\kilo\gram\per\meter\cubed} \\
$\theta$ & turbine azimuthal position, \si{\degree} \\
$\psi$ & phase angle with respect to stream pulsation, \si{\degree} \\
$\kindex{\omega}{H}$ & turbine rotational frequency, \si{\per\second} \\
$\kindex{\omega}{u}$, $\kindex{\omega}{$\alpha$}$, $\kindex{\omega}{h}$ & surging, pitching, heaving frequencies
\end{longtable*}}

\section{Introduction}
\label{sec:intro}
Horizontal-axis wind turbines dominate wind energy production due to their high power efficiency and performance~\cite{Eriksson_2008}.
However, vertical-axis wind turbines can play a significant role as distributed wind energy systems and complement their horizontal-axis counterparts~\cite{Xie:2017dg}.
Some of the main advantages of vertical-axis wind turbines include their:
\begin{inparaenum}[i)]
\item independence on the wind direction~\cite{Buchner_2018},
\item potential to achieve higher power densities~\cite{Dabiri_2011},
\item low levels of sound emission~\cite{Mertens_2003}, and
\item relatively simple mechanics~\cite{Jain_2019}.
\end{inparaenum}
Some key drawbacks of vertical-axis wind turbines have limited their implementation on a large scale.
These drawbacks include a susceptibility to structural fatigue~\cite{Baker_1983}, and inherent aerodynamic complexity related to the occurrence of dynamic stall~\cite{Leishman2002, Veers.2019, Wood.2020}.

Dynamic stall refers to a succession of aerodynamic events that occur when an airfoil exceeds its critical stall angle while undergoing a dynamic motion~\cite{Mccroskey_1981}.
It covers the roll-up of the leading edge shear layer into a large-scale dynamic stall vortex, the separation of the vortex from the blade, the delayed transition to full stall, followed by the increase of the nose-down pitching moment.
The time delay between the moment when the critical stall angle is exceeded and the occurrence of dynamic stall is called the dynamic stall delay and is due to a reaction time delay that decreases with increasing unsteadiness of the motion and a finite vortex formation time~\cite{Deparday.2019,lefouest2021}.
Dynamic stall is also associated with large load hysteresis and delayed flow reattachment.

Dynamic stall is endemic in vertical-axis wind turbines due to the inherent periodic variations in angle of attack experienced by the turbines blades.
The presence and progression of dynamic stall in this context and its influence on the turbine's performance has been studied computationally and experimentally~\cite{Ferreira_2008, Ferreira_2010, Laneville_1986, Buchner_2018}, but few studies have focussed on the individual blade loading.
Yet, to effectively mitigate the negative impacts of dynamic stall and optimize control strategies such as active blade pitching, a detailed understanding and prediction of the individual blades loads, their post-stall fluctuations, and hysteresis is desirable.

Classical analytical unsteady aerodynamic models by \citet{Isaacs_1945,Theodorsen_1949,Greenberg_1947} are typically based on non-stationary incompressible potential-flow theory and have been developed to describe the dynamic loading on wings undergoing various harmonic motions, including surging, pitching, and heaving.
Various studies have analyzed the performance of these classical models through comparison with experimental data of harmonically pitching, plunging, and surging airfoils~\cite{Choi.2015, Granlund2016,Elfering_2020}.
The models typically perform well for low-amplitude motions in absence of flow separation.
But even for high amplitude excursions and in the presence of large scale vortex shedding, these unsteady potential flow models can give valuable insight into the flow response~\cite{Granlund_2014, Strangfeld_2016, Otomo.2021}.

Here, we present a method to adapt \citeauthor{Greenberg_1947}'s potential flow model for coupled pitching and surging flow such that it can be applied to predict the loads on a vertical-axis wind turbine blade.
The model is extended to compute loads on a blade undergoing multi-harmonic oscillations in effective angle of attack and incoming flow velocity by formulating the blade kinematics as a sum of simple harmonic motions.
Each of these functions is a multiple of the main turbine rotational frequency, associated with an individual amplitude, as suggested by \citet{Greenberg_1947}.
The results of the adapted model are compared with experimental data from a scaled-down model of a single-bladed H-type Darrieus wind turbine.

\section{Methods}
\label{sec:methods}

\subsection{Greenberg's original potential flow model}
\label{sec:greenberg_current_app}
\citeauthor{Greenberg_1947}'s potential flow model computes the total forces and moments acting on a blade undergoing a harmonic motion in a pulsating stream.
The model is based on non-stationary incompressible flow theory using the approximation of the wake form by \citet{Theodorsen_1949}.
Greenberg's work was motivated by predicting forces on a helicopter blade in forward flight, which experience harmonic variations in the stream velocity $u(t)$, angle of attack $\alpha (t)$, and vertical displacement or heave $h(t)$ defined as:
\begin{align}
u(t) &= \kindex{u}{0} \left[1+ \sigma \exp{\left(\ii \kindex{\omega}{u}t\right)}\right]
\label{equ:Greenberg_v}\\
\alpha(t) &= \kindex{\alpha}{0} \exp{\left(\ii (\kindex{\omega}{$\alpha$}t + \kindex{\psi}{$\alpha$}\right))}
\label{equ:Greenberg_b}\\
h(t) &= \kindex{h}{0} \exp{\left(\ii (\kindex{\omega}{h}t + \kindex{\psi}{h}\right))},
\label{equ:Greenberg_h}
\end{align}
with $\kindex{u}{0}$ the mean free-stream velocity, and $\kindex{u}{0}\sigma$, $\kindex{\alpha}{0}$, and $\kindex{h}{0}$ the amplitudes of the surging, pitching, and heaving motions.
The phase shifts between the pitching and heaving motions with respect to the surging are given by \kindex{\psi}{$\alpha$} and \kindex{\psi}{h}.
The harmonic frequencies of the velocity, angle of attack, and heaving motions are \kindex{\omega}{u}, \kindex{\omega}{$\alpha$} and \kindex{\omega}{h}.
The associated reduced frequencies are defined as:
\begin{equation}
\kindex{k}{u}= \frac{\kindex{\omega}{u}c}{2\kindex{u}{0}},\quad
\kindex{k}{$\alpha$}= \frac{\kindex{\omega}{$\alpha$}c}{2\kindex{u}{0}},\quad
\kindex{k}{h}= \frac{\kindex{\omega}{h}c}{2\kindex{u}{0}},\quad
\kindex{k}{u+$\alpha$}= \frac{(\kindex{\omega}{u}+\kindex{\omega}{$\alpha$})c}{2\kindex{u}{0}}\quad.
\label{equ:reduced_freq_green}
\end{equation}
Greenberg's final expression for the total force on the blade $P(t)$ is:
\begin{equation}
\begin{split}
P(t) =
-\frac{1}{4}\pi\rho c^{2}\Big[\ddot{h}(t) + u(t)\dot{\alpha}(t) &+\dot{u}(t)(\overline{\alpha}+\alpha(t)) -2d\frac{c}{2}\ddot{\alpha}(t)\Big] \\
-\pi\rho u(t)c \Bigg\{ & \kindex{u}{0}\overline{\alpha} +\sigma \kindex{u}{0} \overline{\alpha} C(\kindex{k}{u}) e^{i \kindex{\omega}{u}t} + \\
&\Big[\frac{c}{2}\left(\frac{1}{2} -2d\right)\dot{\alpha}(t) +\kindex{u}{0}\alpha(t)\Big]C(\kindex{k}{$\alpha$}) \\
& + \dot{h}(t)C(\kindex{k}{h}) + \sigma \kindex{u}{0}\alpha(t) C(\kindex{k}{u+$\alpha$})e^{i \kindex{\omega}{u}t}
\Bigg\},
\end{split}
\label{equ:Greenberg_org}
\end{equation}
with $\dot{\ }$ and $\ddot{\ }$ indicating the first and second time derivative, $\rho$ the fluid density, $\overline{\alpha}$ the time-average or fixed part of the angle of attack, and $d$ the chord-normalised distance between the pitching axis and the center of the airfoil.
If the pitching axis is located at quarter chord, $d=-1/4$.
Theodorsen's complex transfer function $C(k)=F(k)+\ii G(k)$ accounts for the lift attenuation by vorticity shed into the wake and is a function of the reduced frequencies of the motion kinematics.
The real and imaginary parts of Theodorsen's transfer function, $F(k)$ and $G(k)$, are Bessel functions of the first and second kind~\cite{Theodorsen_1949}.

\begin{figure}[tb!]
\centering
\includegraphics{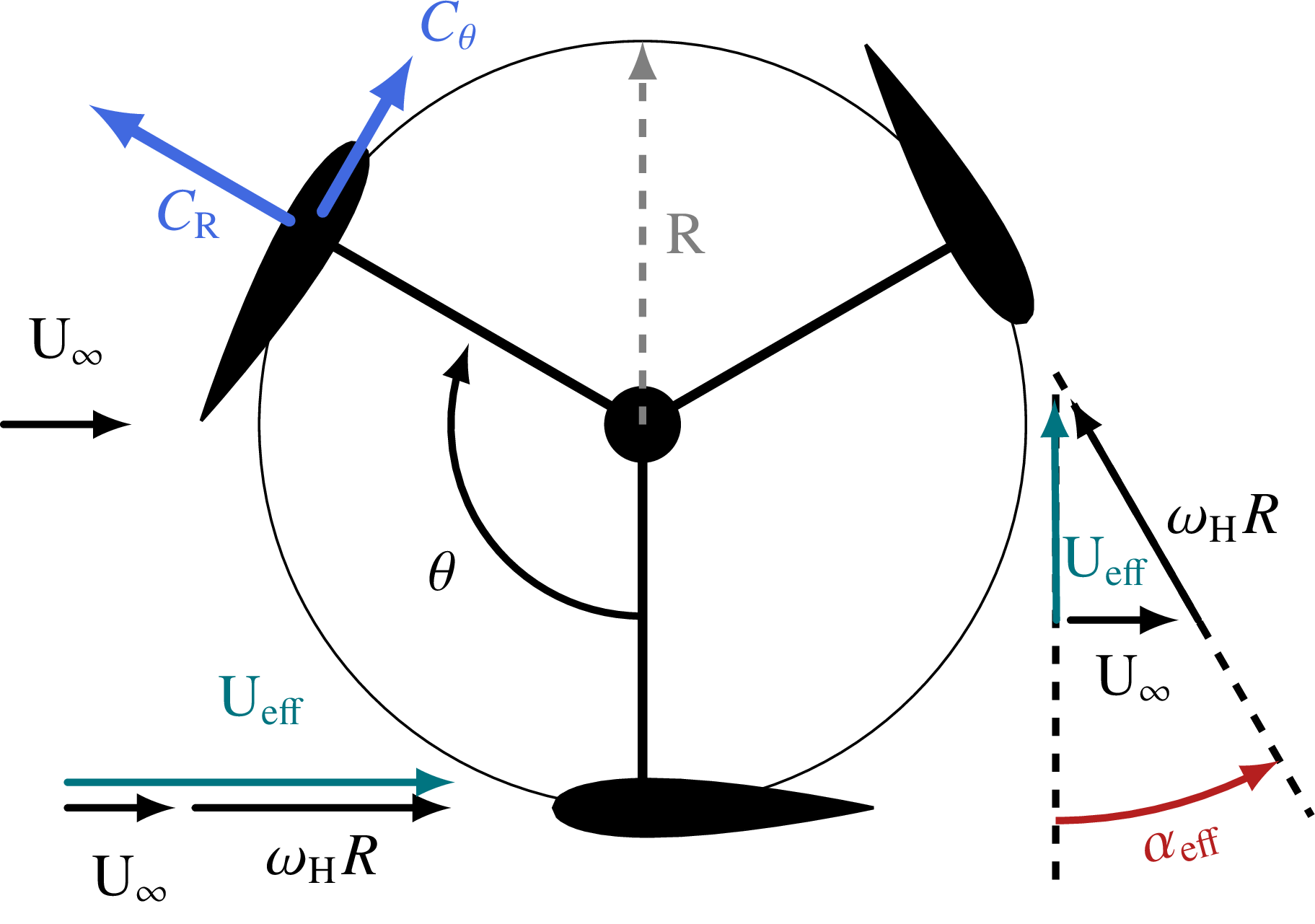}
\caption{Schematic of the definition and orientation of the forces, effective velocity, and angle of attack experienced by the blades in a H-type vertical axis wind turbine.}
\label{fig:profile_VAWT}
\end{figure}

\subsection{Greenberg's model for vertical-axis wind turbine kinematics}
\begin{figure}[tb!]
\centering
\includegraphics{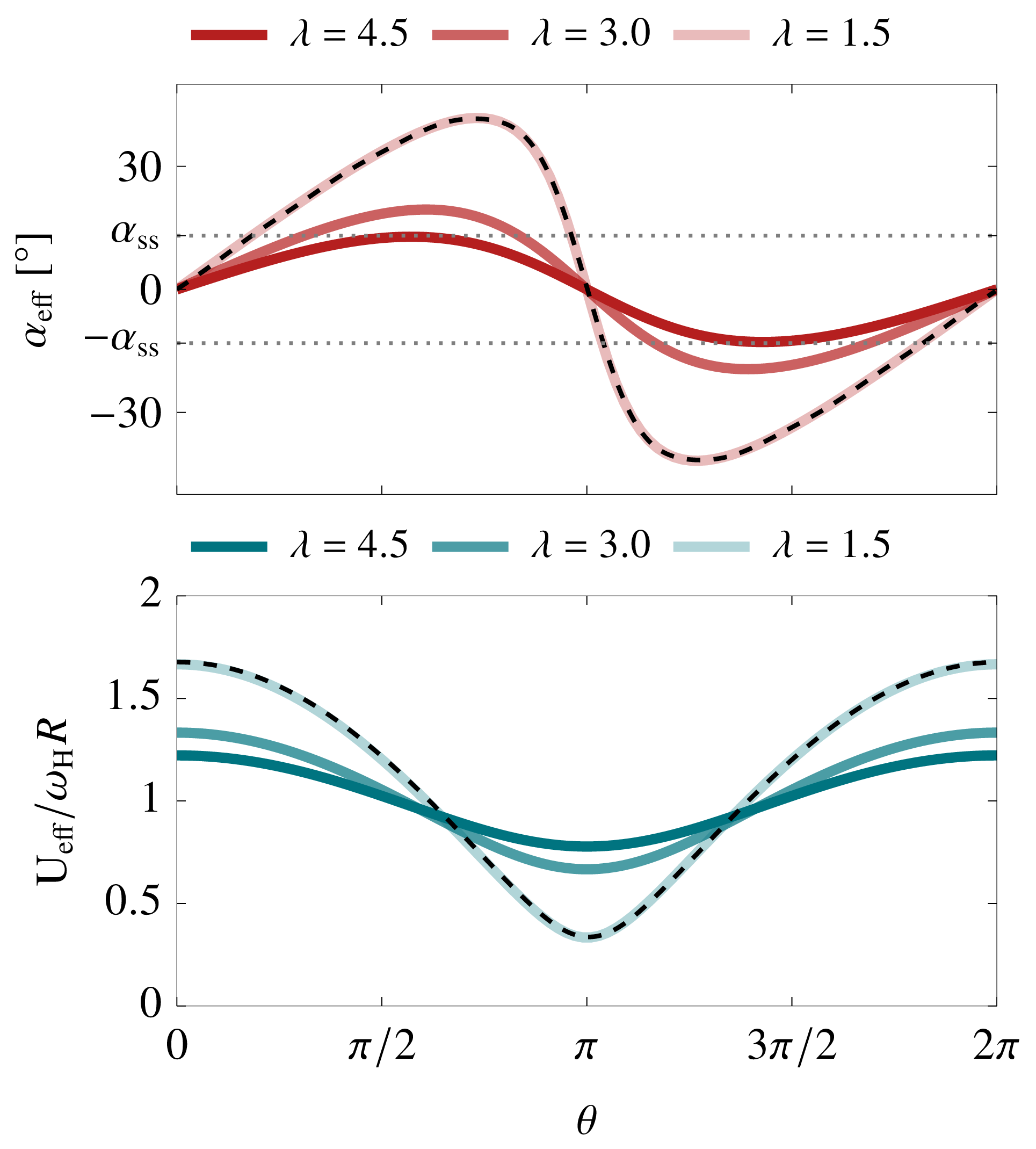}
\caption{Temporal evolution of the a) effective angle of attack and b) effective flow velocity over a single rotation of a vertical-axis wind turbine blade for $\lambda\in \{1.5,3.0,4.5\}$.
The dashed line represents the Fourier series approximation for $\lambda=1.5$
Horizontal dotted lines indicate the static stall angle for a NACA0018 airfoil ($\kindex{\alpha}{ss}=\ang{13.1}$ at $\Rey=\num{50000}$).
}
\label{fig:kinematic}
\end{figure}

The path followed by an H-type Darrieus vertical-axis wind turbine blade is shown in \cref{fig:profile_VAWT}.
The total load acting on the turbine blade can be decomposed into a radial (\kindex{C}{R}) and a tangential (\kindex{C}{$\theta$}) component.
The blade rotates at a constant frequency (\kindex{\omega}{H}) and radius ($R$) around a vertical axis with respect to the free-stream velocity ($\Uinf$).
The ratio between the blade velocity ($\kindex{\omega}{H} R$) and free-stream velocity is known as the tip-speed ratio ($\lambda$) and governs the magnitude of the variation in the effective incoming flow velocity (\Ueff) and effective angle of attack ($\kindex{\alpha}{eff}$) experienced by the blade as a function of the angular coordinate ($\theta$):
\begin{align}
\kindex{\alpha}{eff}(\theta)&= \tan^{-1}\left({\frac{\sin{\theta}}{\lambda+\cos{\theta}}}\right) \label{equ:alphaeff} \\
\Ueff(\theta) &= \Uinf\sqrt{1+2\lambda \cos(\theta) + \lambda^2}\quad.
\label{equ:Ueff}
\end{align}
The kinematic profiles for surging, pitching, and heaving are described by simple harmonic motions in Greenberg's model (\cref{equ:Greenberg_v,equ:Greenberg_b,equ:Greenberg_h}).
This reveals a challenge for the model's application to vertical-axis turbines, where the incoming flow conditions at lower tip-speed ratios deviate from single harmonic oscillations, as highlighted in \cref{fig:kinematic}.
The temporal evolution of the pitch rate and effective flow velocity more closely resemble a saw-tooth shape than a sine wave when the tip-speed ratio decreases and the standard Greenberg model (\cref{equ:Greenberg_org}) would not represent the blade kinematics accurately for low tip-speed ratios.

To use the Greenberg model for the prediction of the forces occurring on a vertical-axis wind turbine blade at low tip-speed ratios, we need to adapt the formulation of the blade kinematics (\cref{equ:Greenberg_v,equ:Greenberg_b}) such that they represent the effective flow variations described by \cref{equ:Ueff,equ:alphaeff}.
As the effective angle of attack starts is minimum and the inflow velocity is maximum at the upwind location indicated by $\theta=0$, we propose to decompose the evolution of the kinematics as Fourier sine and cosine series for the effective angel of attack and flow velocity, respectively.
To facilitate calculations, we write everything in a complex form.
The blade's effective angle of attack is approximated by:
\begin{equation}
\kindex{\alpha}{eff}'(t) = \kindex{a}{0} +  \sum\limits_{n=1}^{N} \kindex{a}{n}\sin(n \kindex{\omega}{H} t) = Re\left[\kindex{a}{0} -\ii \sum\limits_{n=1}^{N} \kindex{a}{n} \exp{\left(\ii n\kindex{\omega}{H} t\right)}\right]
\label{equ:pitching}
\end{equation}
with \kindex{a}{n} the $n^{\rm th}$ Fourier sine coefficient, $\kindex{\omega}{H}$ the rotation frequency of the wind turbine, $Re$ refers to the real part.
The comparison between \cref{equ:pitching} and \cref{equ:alphaeff} reveals that $\kindex{a}{0} = 0$ if there is no pitch angle offset.

In analogy with the formulation of the velocity profile by Greenberg (\cref{equ:Greenberg_v}), we define the approximate effective inflow velocity as:
\begin{equation}
\Ueff'(t)= \Ueffbar\left(1 + \Uosc(t)\right)
\label{equ:surging}
\end{equation}
with $\Ueffbar$ the time-averaged and $\Uosc(t)$ the oscillating part of effective flow velocity.
The latter is written as a Fourier cosine series:
\begin{equation}
\Uosc(t)= \frac{\Ueff'(t)-\Ueffbar}{\Ueffbar} = \kindex{b}{$0$} +  \sum\limits_{n=1}^{N} \kindex{b}{n}\cos(n\kindex{\omega}{H} t)  = Re\left[\kindex{b}{0} + \sum\limits_{n=1}^{N} \kindex{b}{n} \exp{\left(\ii n\kindex{\omega}{H} t\right)}\right]
\label{equ:oscialation}
\end{equation}
with \kindex{b}{n} the $n^{\rm th}$ Fourier cosine coefficient.
For the examples presented here, $N=8$ was found sufficient to accurately fit the effective angle of attack and free-stream velocity variations for $\lambda\geq 1.2$.
The approximations $\kindex{\alpha}{eff}'(t)$ and $\kindex{\U}{eff}'(t)$ for $\lambda=1.5$ are included as examples in \cref{fig:kinematic}.
The difference between the true variation and the Fourier series approximation with $N=8$ terms is so small that it is not visible in the graph.

Using the newly proposed expressions for the effective flow conditions \cref{equ:pitching} and \cref{equ:surging}, we derive the total force prediction according to the Greenberg.
If no active pitching and a constant pitch offset is applied to the blade, we obtain:
\begin{equation}
\begin{split}
P(t) = \frac{1}{4}\pi\rho c^{2}\Big[ \Ueff'(t)&\kindex{\dot{\alpha}'}{eff}(t)  +\kindex{\dot{\U}'}{eff}(t)\kindex{\alpha}{eff}'(t) -2d\frac{c}{2}\kindex{\ddot{\alpha}'}{eff}(t)\Big] \\
-\pi\rho \Ueff'(t) c\Bigg\{ & \Big[\frac{c}{2}\left(\frac{1}{2} -2d\right)\kindex{\dot{\alpha}'}{eff}(t) +\Ueffbar \kindex{\alpha}{eff}'(t)\Big]C(\kindex{k}{$\beta$}) + \\
& \kindex{\alpha}{eff}'(t) C(\kindex{k}{\U+$\alpha$'}) (\Ueff'(t)-\Ueffbar)
\Bigg\}\quad.
\label{equ:Greenberg_VAWT}
\end{split}
\end{equation}
Greenberg's original model includes the functionality for distinct frequencies in the pitching and surging profiles.
For the case of a vertical axis wind turbine, the frequency of the effective angle of attack and inflow velocity variations are the same (\kindex{\omega}{H}) and the relevant reduced frequencies yield:
\begin{equation}
\kindex{k}{\U} = \frac{\kindex{\omega}{H}c}{2\Ueffbar},\quad
\kindex{k}{$\alpha$'} = \frac{\kindex{\omega}{H}c}{2\Ueffbar},\quad
\kindex{k}{\U+$\alpha$'} = \frac{2\kindex{\omega}{H}c}{2\Ueffbar}\quad.
\label{equ:reduced_freq_vawt}
\end{equation}

\begin{figure}[tb!]
\centering
\includegraphics{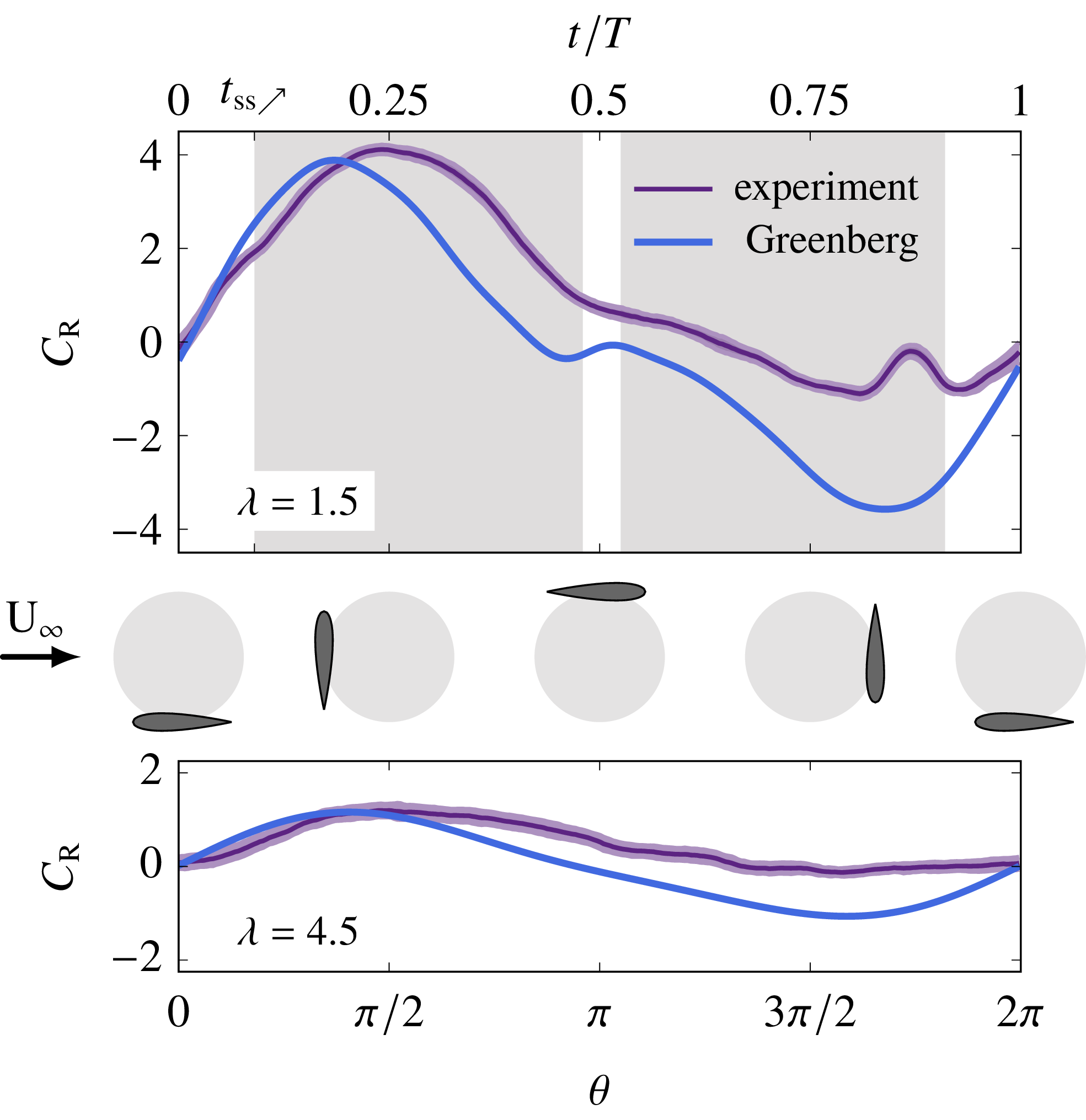}
\caption{Comparison between the experimentally obtained phase-averaged radial force coefficient and the prediction based on the modified Greenberg model for (top) $\lambda=1.5$ and (bottom) $\lambda=4.5$.
The shaded areas in correspond to regions where the magnitude of the effective angle of attack is above the critical stall angle.}
\label{fig:greenberg_comp}
\end{figure}

\section{Comparison with experimental data}
\label{sec:results}
The results from the adapted Greenberg model (\cref{equ:Greenberg_VAWT}) are compared with experimentally obtained phase-averaged radial force coefficient $\kindex{C}{R}$ for $\lambda=1.5$ and $\lambda=4.5$ in \cref{fig:greenberg_comp}. The radial force component is defined in \cref{fig:profile_VAWT} as the projection of the total load orthogonal to the circular path.
The radial component is positive when pointing inward \cref{equ:Greenberg_VAWT}.

The experimental data stems from experiments with a scaled-down model of a motor-driven single-bladed H-type Darrieus wind turbine in a recirculating water channel with a test section of \SI{0.6x0.6x3}{\meter} and maximum flow velocity of $\Uinf$ = \SI{1}{\meter\per\second}~\cite{LeFouest.2022sci}.
The turbine blade has a NACA0018 profile, with chord $c=\SI{6}{\centi\meter}$ and span $s=\SI{15}{\centi\meter}$.
When the blade is orthogonal to the free stream direction, there is a blockage ratio of \SI{2.5}{\percent} relative to the channel cross-section.
The blade has a static stall angle of \ang{13.1} at a Reynolds number of \num{50000}.
The turbine radius was kept constant at \SI{15}{\centi\meter} yielding a chord-to-diameter ratio of \num{0.2}.
The pitching axis of the blade is set at the quarter-chord such that $d= -1/4$.
Experiments are conducted at a Reynolds number \num{50000} based on the blade rotational velocity of \SI{0.89}{\hertz}.
The unsteady blade loads were measured by twenty strain gauges forming five full Wheatstone bridge channels on the blade shaft.
The strain guages measure the tangential (\kindex{C}{$\theta$}) and orthogonal (\kindex{C}{R}) shear components illustrated in \cref{fig:profile_VAWT}, and the pure bending moments and axial torque.
The load measurement system was calibrated in-situ yielding a calibration matrix with a \SI{95}{\percent} confidence interval of the load coefficients.
The inertial centripetal force of the turbine was quantified by operating the turbine in air and the results was removed from the measured signal to obtain the aerodynamic load responses that are presented here.
Aerodynamic load responses are phase-averaged over \num{100} cycles and are conditioned using a low-pass filter with cut-off frequency of \SI{30}{\hertz}.

The shaded areas in \cref{fig:greenberg_comp}(top) correspond to the regions where the magnitude of the effective angle of attack is above the critical stall angle.
The phase-averaged experimental data is presented by the solid line and surrounded by a lighter band to indicate the standard deviation due to cycle-to-cycle variations.
The results are compared with the predictions of the Greenberg model obtained by projecting the result of \cref{equ:Greenberg_VAWT} in the radial direction.
Overall, the Greenberg model predicts the radial forces well at the start of the cycle, when the blade is in the windward position, and the maximum force coefficient during the upwind portion of the cycle ($0<\theta<\pi$) even for the low tip speed ratio case where dynamic stall occurs.
For the low tip speed case, the Greenberg model deviates from the experimental results once the effective angle of attack exceeds the critical stall angle.
The maximum value of the radial force coefficient is still well predicted, but its timing is not.
This delay in radial load response at low tip-speed ratio can be attributed to the formation and separation of a leading-edge dynamic stall vortex, consistent with flow observations documented by ~\citet{Buchner_2018}.
The post dynamic stall load response remains low during almost the entire downwind portion of the cycle ($\pi<\theta<2\pi$) as a result of the massive flow separation and blade interaction with the shed vortex in the upwind portion \citep{Ferreira_2008}, neither of which can not be taken into account by Greenberg's model.

For higher tip speed ratios, the maximum radial load coefficient and its timing are well predicted by the Greenberg model.
Again, the model fails during the downwind portion of the cycle even though the magnitude of the effective angle of attack remains below the static stall angle (\ang{13}) for $\lambda=4.5$.
During the downwind part of the cycle, the magnitude of the experimental radial force coefficient is lower than predicted by the Greenberg model.
During this part of the rotation, the blade interacts with its wake \citep{ferrer2015blade} which prevents the build up of radial force.
The blade wake interaction is not taken into account in the Greenberg model.

\section{Conclusion}
\label{sec:conclusion}

We have derived a solution for Greenberg's unsteady aerodynamics model to predict the inviscid aerodynamic loads acting on an H-type Darrieus wind turbine blade.
The turbine blade's inherent variation in effective angle of attack and effective flow velocity are reconstructed using Fourier sine and cosine series such that they fit the format required by the Greenberg model.
The comparison between the predictions by the Greenberg model and experimentally obtained phase-averaged radial force evolutions show that the inviscid Greenberg model predicts well the loads at the start of the upwind portion and the maximum loads during upwind, but fails during the downwind portion when flow separation and blade wake interaction occur.

The proposed application of Greenberg's model to vertical-axis wind turbine kinematics show a great potential to diagnose regions of separated flow and for quantifying the relative influences of dynamic stall and intrinsic turbine kinematics on the blade loading.
Future research can readily extend this method to any airfoil undergoing an arbitrary combination of pitching, surging and heaving, following a kinematic profile which can be approximated by a Fourier series.

\section*{Acknowledgments}
This work was supported by the Swiss national science foundation under grant number PYAPP2\_173652.

\bibliography{TN}

\begin{thebibliography}{27}
\newcommand{\enquote}[1]{``#1''}
\providecommand{\natexlab}[1]{#1}
\providecommand{\url}[1]{\texttt{#1}}
\providecommand{\urlprefix}{URL }
\expandafter\ifx\csname urlstyle\endcsname\relax
  \providecommand{\doi}[1]{\discretionary{}{}{}https://doi.org/#1}\else
  \providecommand{\doi}[1]{\discretionary{}{}{}\urlstyle{rm}\url{https://doi.org/#1}}\fi

\bibitem[{Eriksson et~al.(2008)Eriksson, Bernhoff, and Leijon}]{Eriksson_2008}
Eriksson, S., Bernhoff, H., and Leijon, M., \enquote{Evaluation of different
  turbine concepts for wind power,} \emph{Renewable and Sustainable Energy
  Reviews}, Vol.~12, No.~5, 2008, pp. 1419--1434.
\newblock \doi{10.1016/j.rser.2006.05.017}.

\bibitem[{Xie et~al.(2017)Xie, Archer, Ghaisas, and Meneveau}]{Xie:2017dg}
Xie, S., Archer, C.~L., Ghaisas, N., and Meneveau, C., \enquote{{Benefits of
  collocating vertical‐axis and horizontal‐axis wind turbines in large wind
  farms},} \emph{Wind Energy}, Vol.~20, No.~1, 2017, p. 45 62.
\newblock \doi{10.1002/we.1990},
  \urlprefix\url{https://onlinelibrary.wiley.com/doi/full/10.1002/we.1990}.

\bibitem[{Buchner et~al.(2018)Buchner, Soria, Honnery, and
  Smits}]{Buchner_2018}
Buchner, A., Soria, J., Honnery, D., and Smits, A., \enquote{Dynamic stall in
  vertical axis wind turbines:scaling and topological considerations,}
  \emph{Journal of Fluid Mechanics}, Vol. 841, 2018, pp. 746--766.
\newblock \doi{10.1017/jfm.2018.112}.

\bibitem[{Dabiri(2011)}]{Dabiri_2011}
Dabiri, J., \enquote{Potential order-of-magnitude enhancement of wind farm
  power density via counter-rotating vertical-axis wind turbine arrays,}
  \emph{Journal of Renewable and Sustainable Energy}, Vol.~3, No.~4, 2011, p.
  043104.
\newblock \doi{10.1063/1.3608170}.

\bibitem[{Mertens et~al.(2003)Mertens, van Kuik, and van Bussel}]{Mertens_2003}
Mertens, S., van Kuik, G., and van Bussel, G., \enquote{Performance of an
  H-Darrieus in the Skewed Flow on a Roof,} \emph{Journal of Solar Energy
  Engineering}, Vol. 125, No.~4, 2003, pp. 433--440.
\newblock \doi{10.1115/1.1629309}.

\bibitem[{Jain and Saha(2019)}]{Jain_2019}
Jain, S., and Saha, U., \enquote{The State-of-the-Art Technology of H-Type
  Darrieus Wind Turbine Rotors,} \emph{Journal of Energy Resources Technology},
  Vol. 142, No.~3, 2019.
\newblock \doi{10.1115/1.4044559}.

\bibitem[{Baker(1983)}]{Baker_1983}
Baker, J.~R., \enquote{Features to aid or enable self starting of fixed pitch
  low solidity vertical axis wind turbines,} \emph{Journal of Wind Engineering
  and Industrial Aerodynamics}, Vol.~15, No. 1-3, 1983, pp. 369--380.
\newblock \doi{10.1016/0167-6105(83)90206-4}.

\bibitem[{Leishman(2002)}]{Leishman2002}
Leishman, J.~G., \enquote{{Challenges in modelling the unsteady aerodynamics of
  wind turbines},} \emph{Wind Energy}, Vol.~5, No. 2-3, 2002, p. 85 132.
\newblock \doi{10.1002/we.62},
  \urlprefix\url{http://doi.wiley.com/10.1002/we.62}.

\bibitem[{Veers et~al.(2019)Veers, Dykes, Lantz, Barth, Bottasso, Carlson,
  Clifton, Green, Green, Holttinen, Laird, Lehtomäki, Lundquist, Manwell,
  Marquis, Meneveau, Moriarty, Munduate, Muskulus, Naughton, Pao, Paquette,
  Peinke, Robertson, Rodrigo, Sempreviva, Smith, Tuohy, and Wiser}]{Veers.2019}
Veers, P., Dykes, K., Lantz, E., Barth, S., Bottasso, C.~L., Carlson, O.,
  Clifton, A., Green, J., Green, P., Holttinen, H., Laird, D., Lehtomäki, V.,
  Lundquist, J.~K., Manwell, J., Marquis, M., Meneveau, C., Moriarty, P.,
  Munduate, X., Muskulus, M., Naughton, J., Pao, L., Paquette, J., Peinke, J.,
  Robertson, A., Rodrigo, J.~S., Sempreviva, A.~M., Smith, J.~C., Tuohy, A.,
  and Wiser, R., \enquote{{Grand challenges in the science of wind energy.}}
  \emph{Science}, Vol. 366, No. 6464, 2019.
\newblock \doi{10.1126/science.aau2027}.

\bibitem[{Wood(2020)}]{Wood.2020}
Wood, D., \enquote{{Grand Challenges in Wind Energy Research},} \emph{Frontiers
  in Energy Research}, Vol.~8, 2020, p. 624646.
\newblock \doi{10.3389/fenrg.2020.624646}.

\bibitem[{Mccroskey(1981)}]{Mccroskey_1981}
Mccroskey, W.~J., \enquote{The Phenomenon of Dynamic Stall,} Tech. rep.,
  National Aeronautics and Space Administration, 1981.

\bibitem[{Deparday and Mulleners(2019)}]{Deparday.2019}
Deparday, J., and Mulleners, K., \enquote{{Modeling the interplay between the
  shear layer and leading edge suction during dynamic stall},} \emph{Physics of
  Fluids}, Vol.~31, No.~10, 2019, p. 107104.
\newblock \doi{10.1063/1.5121312}.

\bibitem[{Fouest et~al.(2021)Fouest, Deparday, and Mulleners}]{lefouest2021}
Fouest, S.~L., Deparday, J., and Mulleners, K., \enquote{{The dynamics and
  timescales of static stall},} \emph{Journal of Fluids and Structures}, Vol.
  104, No. 103304, 2021, pp. 1---11.
\newblock \doi{10.1016/j.jfluidstructs.2021.103304}.

\bibitem[{Simao~Ferreira et~al.(2008)Simao~Ferreira, van Kuik, van Bussel, and
  Scarano}]{Ferreira_2008}
Simao~Ferreira, C., van Kuik, G., van Bussel, G., and Scarano, F.,
  \enquote{Visualization by PIV of dynamics stall on a vertical axis wind
  turbine,} \emph{Experiments in Fluids}, Vol.~46, No.~1, 2008, pp. 97--108.
\newblock \doi{10.1007/s00348-008-0543-z}.

\bibitem[{Ferreira et~al.(2010)Ferreira, van Zuijlen, Bijl, van Bussel, and van
  Kuik}]{Ferreira_2010}
Ferreira, C., van Zuijlen, A., Bijl, H., van Bussel, G., and van Kuik, G.,
  \enquote{Simulating dynamic stall in a two-dimensional vertical-axis wind
  turbine: verification and validation with particle image velocimetry data,}
  \emph{Wind Energy}, Vol.~13, No.~1, 2010, pp. 1--17.
\newblock \doi{10.1002/we.330}.

\bibitem[{Laneville and Vittecoq(1986)}]{Laneville_1986}
Laneville, A., and Vittecoq, P., \enquote{Dynamic Stall: The Case of the
  Vertical Axis Wind Turbine,} \emph{Journal of Solar Energy Engineering}, Vol.
  108, No.~2, 1986, pp. 140--145.
\newblock \doi{10.1115/1.3268081}.

\bibitem[{Isaacs(1945)}]{Isaacs_1945}
Isaacs, R., \enquote{Airfoil Theory for Flows of Variable Velocity,}
  \emph{Journal of the Aeronautical Sciences}, Vol. 165, 1945, pp. 113--117.
\newblock \doi{10.2514/8.11202}.

\bibitem[{Theodorsen(1935)}]{Theodorsen_1949}
Theodorsen, T., \enquote{General Theory of an aerodynamic instability and the
  mechanism of flutter,} Tech. Rep. 496, Langley Memorial Aeronautical
  Laboratory, 1935.

\bibitem[{Greenberg(1947)}]{Greenberg_1947}
Greenberg, J., \enquote{Airfoil in sinusoidal motion in a pulsating stream.
  NACA Tech,} Tech. rep., Note 1326, 1947.

\bibitem[{Choi et~al.(2015)Choi, Colonius, and Williams}]{Choi.2015}
Choi, J., Colonius, T., and Williams, D.~R., \enquote{{Surging and plunging
  oscillations of an airfoil at low Reynolds number},} \emph{Journal of Fluid
  Mechanics}, Vol. 763, 2015, pp. 237--253.
\newblock \doi{10.1017/jfm.2014.674}.

\bibitem[{Granlund et~al.(2016)Granlund, Ol, and Jones}]{Granlund2016}
Granlund, K.~O., Ol, M.~V., and Jones, A.~R., \enquote{{Streamwise oscillation
  of airfoils into reverse flow},} \emph{AIAA Journal}, Vol.~54, No.~5, 2016,
  pp. 1628--1636.
\newblock \doi{10.2514/1.j054674}.

\bibitem[{Elfering and Granlund(2020)}]{Elfering_2020}
Elfering, K., and Granlund, K., \enquote{Lift Equivalence and Cancellation for
  Airfoil Surge–Pitch–Plunge Oscillations,} \emph{AIAA Journal}, Vol.~58,
  2020.
\newblock \doi{https://doi.org/10.2514/1.J059068}.

\bibitem[{Granlund et~al.(2014)Granlund, Monnier, Ol, and
  Williams}]{Granlund_2014}
Granlund, K., Monnier, B., Ol, M., and Williams, D., \enquote{Airfoil
  longitudinal gust response in separated vs. attached flows,} \emph{Physics of
  Fluids}, Vol.~26, 2014.
\newblock \doi{https://doi.org/10.1063/1.4864338}.

\bibitem[{Strangfeld et~al.(2016)Strangfeld, Müller-Vahl, Nayeri, Paschereit,
  and Greenblatt}]{Strangfeld_2016}
Strangfeld, C., Müller-Vahl, H., Nayeri, C., Paschereit, C., and Greenblatt,
  D., \enquote{Airfoil in a high amplitude oscillating stream,} \emph{Journal
  of Fluid Mechanics}, Vol. 793, 2016, pp. 79--108.
\newblock \doi{doi:10.1017/jfm.2016.126}.

\bibitem[{Ōtomo et~al.(2021)Ōtomo, Henne, Mulleners, Ramesh, and
  Viola}]{Otomo.2021}
Ōtomo, S., Henne, S., Mulleners, K., Ramesh, K., and Viola, I.~M.,
  \enquote{{Unsteady lift on a high-amplitude pitching aerofoil},}
  \emph{Experiments in Fluids}, Vol.~62, No.~6, 2021, pp. 1--18.
\newblock \doi{10.1007/s00348-020-03095-2}.

\bibitem[{Le~Fouest et~al.(2022)Le~Fouest, Bensason, and
  Mulleners}]{LeFouest.2022sci}
Le~Fouest, S., Bensason, D., and Mulleners, K., \enquote{Asymmetry of
  timescales, loads, and flow structures on a vertical-axis wind turbine
  blade,} \emph{AIAA Scitech forum}, San Diego, USA, 2022, pp. 1--12.

\bibitem[{Ferrer and Willden(2015)}]{ferrer2015blade}
Ferrer, E., and Willden, R.~H., \enquote{Blade--wake interactions in cross-flow
  turbines,} \emph{International Journal of Marine Energy}, Vol.~11, 2015, pp.
  71--83.

\end{thebibliography}
\end{document}